\documentclass[aps,prl,superscriptaddress,twocolumn,amsfonts,amsmath,a4paper,showpacs]{revtex4}
\usepackage{graphicx}
\usepackage{multirow}
\usepackage{booktabs}

\begin{document}

\title{Polyelectrolyte-Compression Forces between Spherical DNA Brushes}

\author{Kati~Kegler}
\affiliation{Institute for Experimental Physics I, University of
Leipzig, Lin{\'e}estra{\ss}e 5, D-04103 Leipzig, Germany}

\author{Martin~Konieczny}
\affiliation{Institute for Theoretical Physics II: Soft Matter,
Heinrich-Heine-Universit\"{a}t D\"{u}sseldorf,
Universit{\"a}tsstra{\ss}e 1,
D-40225 D{\"u}sseldorf, Germany}

\author{Gustavo~Dominguez-Espinosa}
\affiliation{Institute for Experimental Physics I, University of
Leipzig, Lin{\'e}estra{\ss}e 5, D-04103 Leipzig, Germany}

\author{Christof~Gutsche}
\affiliation{Institute for Experimental Physics I, University of
Leipzig, Lin{\'e}estra{\ss}e 5, D-04103 Leipzig, Germany}

\author{Matthias~Salomo}
\affiliation{Institute for Experimental Physics I, University of
Leipzig, Lin{\'e}estra{\ss}e 5, D-04103 Leipzig, Germany}

\author{Friedrich~Kremer}
\affiliation{Institute for Experimental Physics I, University of
Leipzig, Lin{\'e}estra{\ss}e 5, D-04103 Leipzig, Germany}

\author{Christos~N.~Likos}
\affiliation{Institute for Theoretical Physics II: Soft Matter,
Heinrich-Heine-Universit\"{a}t D\"{u}sseldorf,
Universit{\"a}tsstra{\ss}e 1,
D-40225 D{\"u}sseldorf, Germany}

\affiliation{The Erwin Schr{\"o}dinger International Institute for
Mathematical Physics, Boltzmanngasse 9, A-1090 Vienna, Austria}

\begin{abstract}
Optical tweezers are employed to measure the forces of interaction
within {\it a single pair} of DNA-grafted colloids
in dependence of the molecular weight of the DNA-chains, and the 
concentration and valence of the surrounding ionic medium. The 
resulting forces are short-range and set in as the surface-to-surface 
distance between the colloidal cores reaches the value of the brush 
height. The measured force-distance dependence is analyzed by means 
of a theoretical treatment based on the compression of the chains on 
the surface of the opposite-lying colloid. Quantitative agreement with 
the experiment is obtained for all parameter combinations.
\end{abstract}

\date{\today}

\maketitle

Surface treatment of colloidal particles
and the ensuing manipulation and control of their interaction
properties 
is a topic of high and lasting interest, on the grounds of both 
technological relevance and fundamental importance. On the first
count, the main issue pertains to the fact that surface treatment
is necessary to achieve colloidal stabilization by inducing thereby
a repulsive force that acts against the ubiquitous dispersion 
attractions between the colloids. Charge stabilization and steric
stabilization, the latter being caused by grafted polymer chains,
are the two most common mechanisms, whereas grafting of polyelectrolyte
(PE) chains on a colloid provides a natural combination of both and
results to an electrosteric repulsion. On the second count, surface
treatment by polymer grafting provides the possibility to tune
the effective colloid interaction by `dressing' the hard sphere 
potential with a soft tail, whose range, strength and overall
functional form can be controlled by changing the properties of
the polymer brush, e.g., its grafting density, height or charge.
Systems interacting by a combination of a hard sphere potential
and a subsequent short-range repulsion show a tremendous variety in
their equilibrium \cite{bolhuis:prl:94,jagla:99,norizoe:epl:05,glaser:epl:07}
and dynamical \cite{pablo:pre:06, emanuela:pre:05, largo:pre:07} properties. 

Considerable work has been carried out in the study of the so-called
{\it osmotic} PE-brushes \cite{pincus:mm:1, borisov:EPJB:1, mei:prl:06},
which result for high surface grafting densities and are characterized 
by
the fact that they spherically condense the vast majority of the counterions 
released by the chains. These, in turn, bring about an entropic effective
force between the brushes, which has been quantitatively
analyzed for PE-brushes \cite{jusufi:cps:04} 
and stars \cite{jusufi:prl:02}.
On the other hand, little
is known for the opposite case of low surface grafting density,
for which the theoretical considerations that lead to the 
interaction between osmotic
brushes break down. In this Letter, we investigate by a combination
of sensitive and accurate experiments and theoretical analysis the
effective forces between spherical DNA brushes and establish 
a novel mechanism of interaction between those, which results
from the mutual compression of PE-chains of the colloids against
the surface of each other. The quantitative characteristics of the
resulting forces are vastly different from those between osmotic
brushes.

The experimental investigation was based on the measurement of
the force-distance dependence between the brushes employing
optical tweezers.
Optical tweezers offer
unprecedented accuracy down to the pN-domain and $3\,{\rm nm}$ in measuring
forces and position, respectively. By monitoring the force-distance
dependencies between two grafted colloids it is possible to know how
the different physicochemical properties (molecular weight, grafting
density, ionic strength of the surrounding medium) affect 
the effective interaction between the grafted colloids.
The force $F(D)$ and the surface-to-surface separation $D$ between
two identical, negatively charged DNA-grafted colloids is measured
using optical tweezers in a identical set-up as used in Ref.
\cite{kegler:prl:07}.

\begin{figure}
  \includegraphics[width=6.0cm]{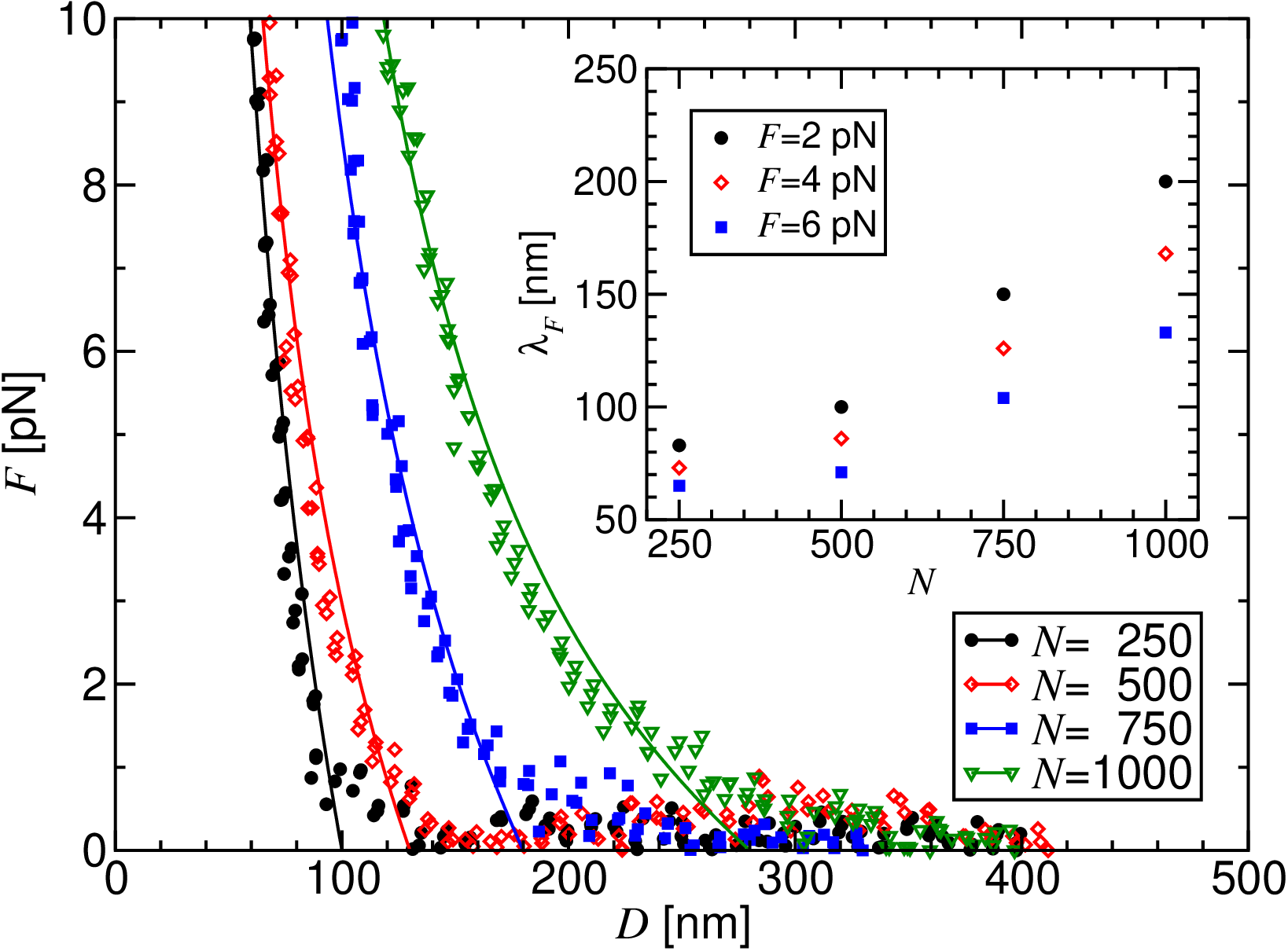}
  \caption{
  $F(D)$ curves between DNA-grafted colloids in buffered solution 
  (10 mM C$_{4}$H$_{11}$NO$_{3}$, pH 8.5) for grafting density 
  $\sigma=8.2\cdot 10^{-5}$~nm$^{-2}$ and various bp-number $N$,
  indicated in the legend. Symbols: experiments; lines: theory according to
  Eq.~(\ref{vcompr:eq}), with the values for $L_0$ and $Z_{\rm eff}$ given 
  in Table \ref{table:tab}. The inset illustrates the dependence of the 
  interaction length $\lambda_F$ on $N$ for three different values of the 
  force.}
\label{bp:fig}
\end{figure}

We employed colloidal particles with a hard core radius 
$R_{\rm c} = 1100\,{\rm nm}$, on which DNA-strands with various
numbers of base pairs (bp) and grafting densities $\sigma$ were
chemically anchored. The brush has only a slight deviation from
planarity, which allows us to relate also to known facts from
planar PE-brushes in what follows.
The force separation dependence between DNA-grafted
colloids with $\sigma = 8.2\cdot 10^{-5}\,{\rm chains/nm^2}$
and varying molecular weights of the chains 
is shown in Fig.~\ref{bp:fig}. By using shorter and shorter DNA-segments,
the force displays a gradual transition from a soft to a hard sphere 
potential.  The theoretical curves based on the model explained below are 
also incorporated. 
A reliable estimate for the brush thickness is deduced 
by determining the interaction length $\lambda_F$ at forces of 
$2\,{\rm pN}$, $4\,{\rm pN}$, and $6\,{\rm pN}$, see inset of 
Fig.~\ref{bp:fig}.
For molecular weights between 500 bp and 1000 bp, a linear scaling is 
observed, in accordance with Ref.~\cite{borisov:EPJB:1}. Deviations at 
chain length 250 bp are attributed to the relative increase of the 
interaction forces between uncoated surfaces. 

\begin{figure}
  \includegraphics[width=6.0cm]{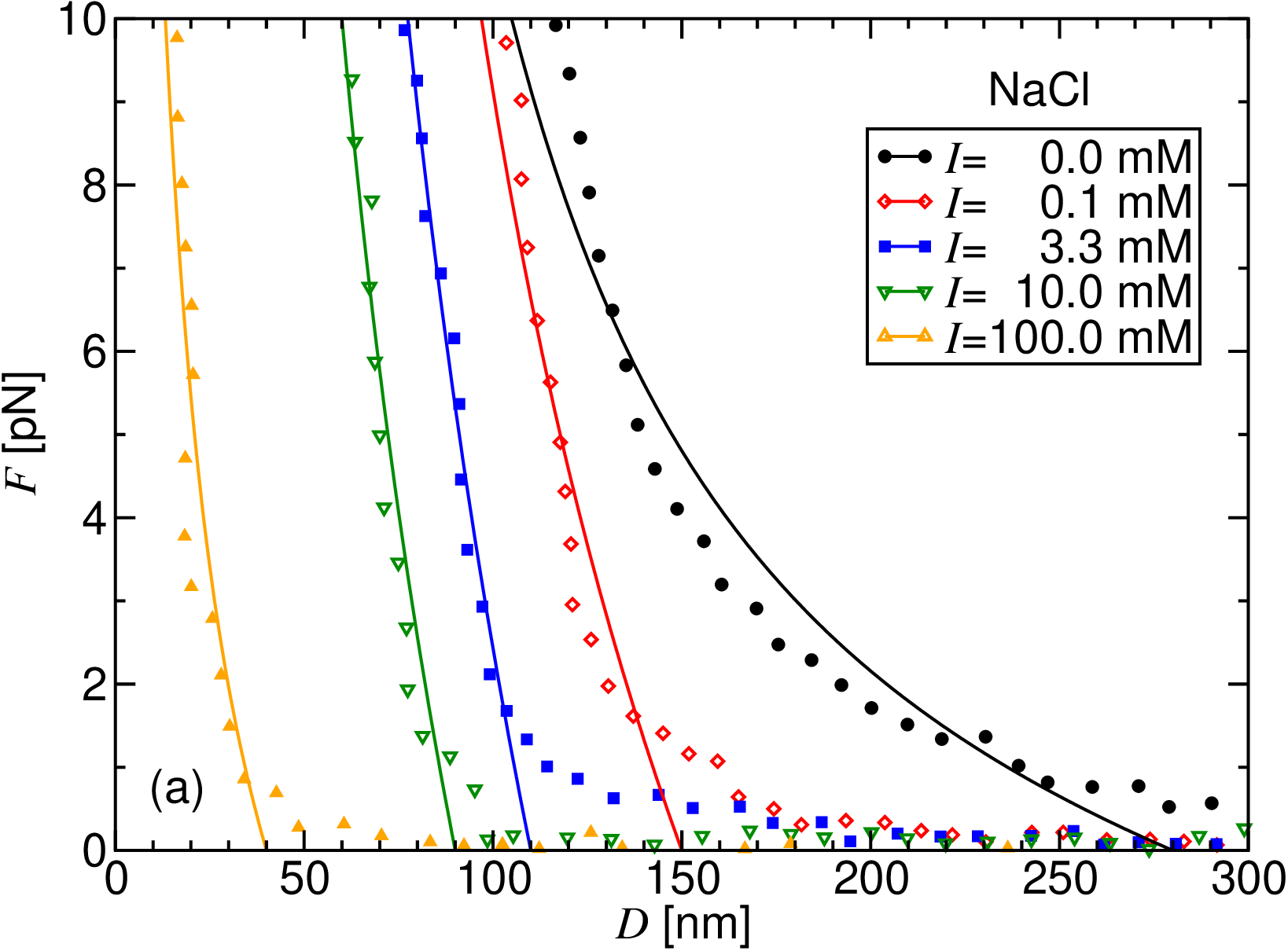}
  \includegraphics[width=6.0cm]{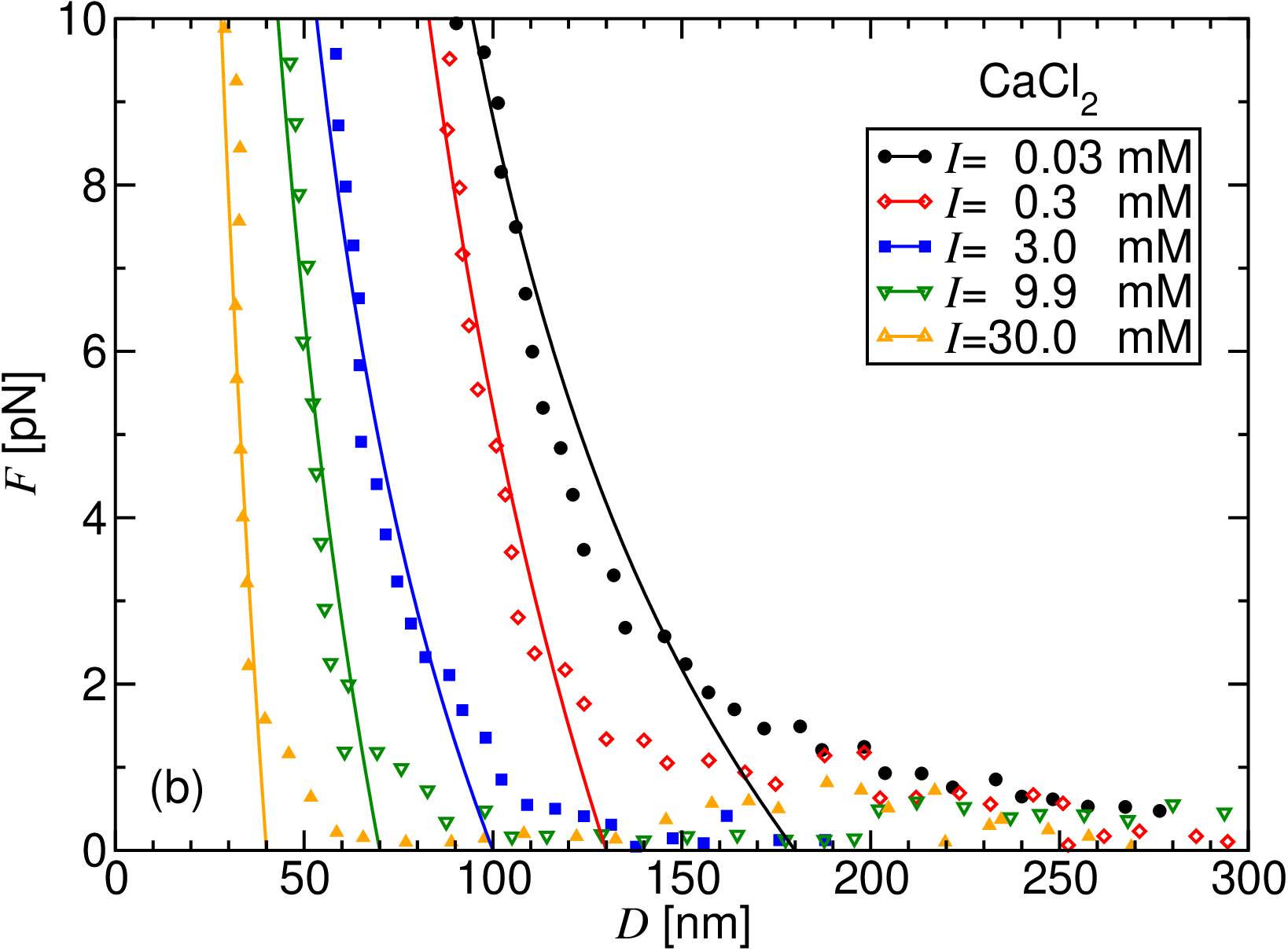}
  \includegraphics[width=6.0cm]{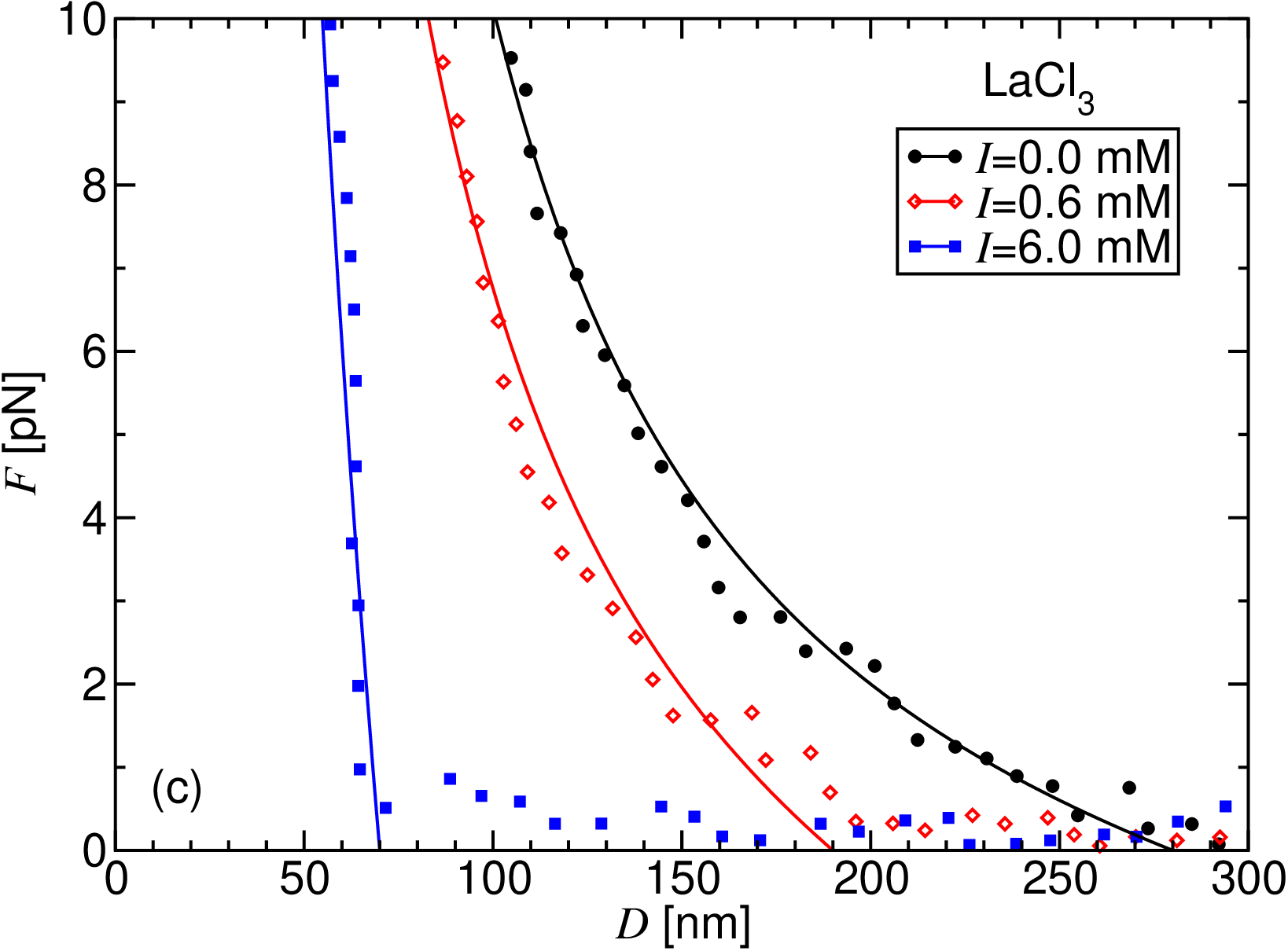}
  \caption{
  $F(D)$ dependence for various ionic strengths $I$ and types of salt.
  (a) NaCl; (b) CaCl$_2$; (c) LaCl$_3$. Here,
  $N=1000$ bp and 
  $\sigma=8.2\cdot 10^{-5}$~nm$^{-2}$.
  Symbols: experiment; lines: theory according to Eq.~(\ref{vcompr:eq}),
  $L_0$ and $Z_{\rm eff}$ are given in 
  Table \ref{table:tab}.}
  \label{salt:fig}
\end{figure}

\begin{figure}
  \includegraphics[width=6.0cm]{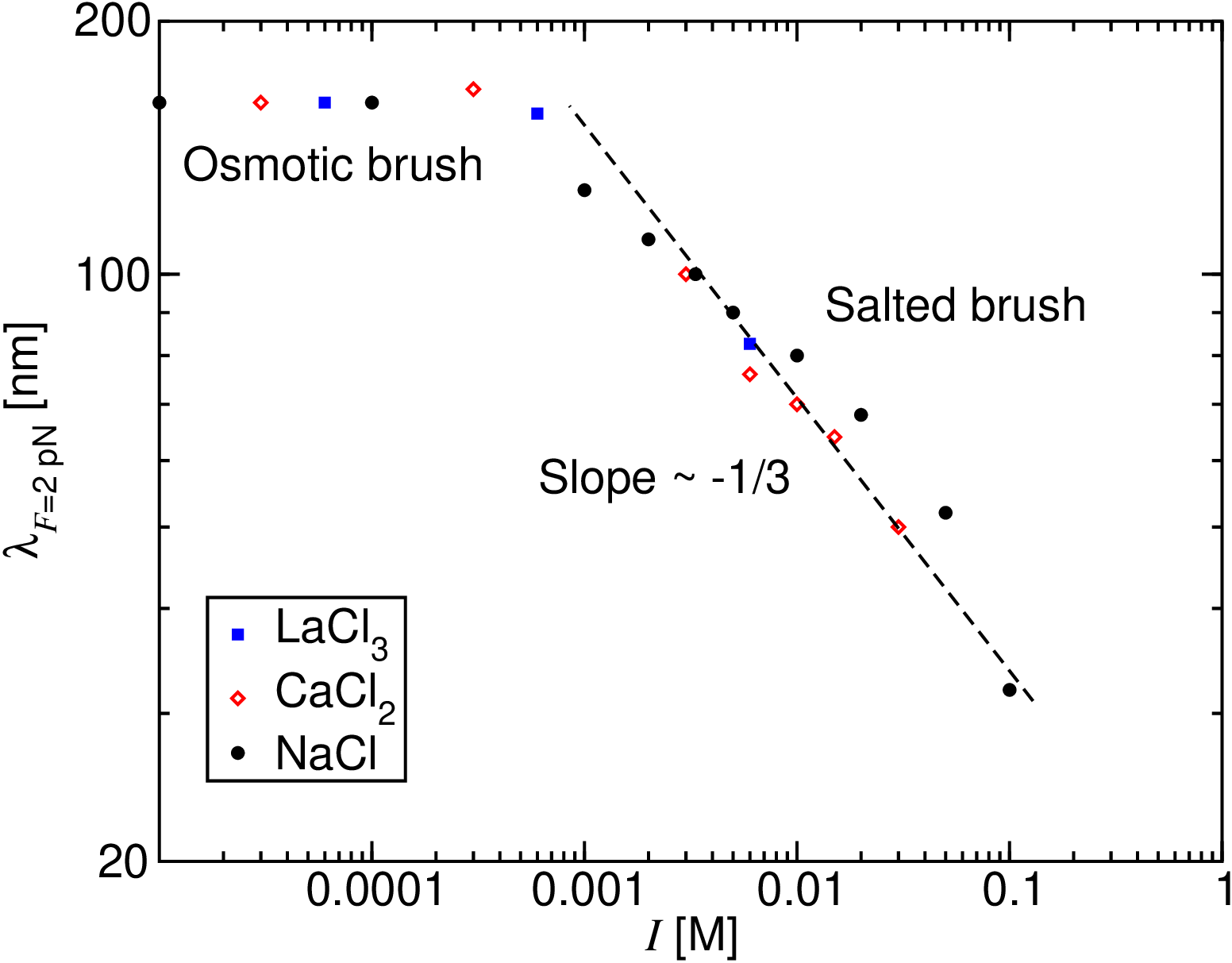}\\
  \caption{
  Double-logarithmic plot of the interaction length for a force $F=2$ pN 
  versus the ionic strength of the added salt. Here, the molecular weight 
  of the grafted DNA is $N=1000$. Different types of symbols correspond 
  to different salt valencies. The dashed line of slope $-1/3$ indicates 
  theoretical scaling law predictions for comparison.}
\label{scale:fig}
\end{figure}

In Fig.~\ref{salt:fig} we show the $F(D)$ dependence on the 
concentration and valency of added salt, using the ionic strength
$I = (\sum_i c_i z_i^2)/2$ as a parameter, where the sum is carried
over all salt ions of concentration $c_i$ and valency $z_i$.
In order to eliminate possible uncertainties due to variations among 
the colloids, the experiments for the data shown in Fig.~\ref{salt:fig} 
were carried out with 
one single pair of colloids
for which the solvent is exchanged. With increasing salt concentration the 
force-separation dependence becomes shorter-range, reflecting the 
transition from an osmotic to a salted brush \cite{kegler:prl:07} and the 
concomitant shrinkage of the latter. The trends are the same independently
of the counterion valency (NaCl, ${\rm CaCl_2}$, and ${\rm LaCl_3}$).
As for the salt-free case, we obtain an estimate for the brush
thickness as the interaction length $\lambda_F$ at the force of 
$2\,{\rm pN}$; results for this quantity are shown in 
Fig.~\ref{scale:fig}.
The slope of the brush thickness versus ionic strength is close to
$0.3\pm 0.05$, in good agreement with the scaling law
\cite{pincus:mm:1}. The transition from the osmotic to the salted
brush takes place when the external salt content equals the
counterion concentration inside the brush. 
A contact between the solid surfaces of the particles
is not observed at low salt concentration ($ < 1\,{\rm mM}$ NaCl)
and high forces (up to $100\,{\rm pN}$) in accordance with known results
for planar brushes \cite{rabin:91}.

In order to gain a deeper understanding of the underlying
physical mechanisms that give rise to the measured forces, we
rely on recent theoretical and simulational work on star-branched
polyelectrolytes and the related system of spherical polymer
brushes \cite{jusufi:prl:02,jusufi:cps:04,konieczny:jcp:06}.
It has been shown \cite{jusufi:prl:02} 
that the main physical mechanism giving rise to a
(soft) repulsion between star-shaped polyelectrolytes comes from
the entropic contribution of the counterions that are spherically
trapped within the star's corona. This consideration has been
extended to spherical brushes, which possess a rigid, colloidal core.
An analytical expression for the entropic effective
interaction, $V_{\rm en}(D)$, between two brushes at
surface-to-surface distance $D$ has been derived,
reading \cite{jusufi:cps:04}:
\begin{align}
\beta V_{\rm en}&(D)=N_{\rm trap}
\left\{
\frac{D+2R_{\rm c}}{2RK}
\ln^2\left(\frac{D+2R_{\rm c}}{2R}\right)\right.\nonumber\\
&\left.+2R_{\rm c}\left[\frac{2}{RK}-\frac{1}{L_0}\right]\ln\left(\frac{R_{\rm c}}{R}\right)
+\ln\left(\frac{2L_0}{RK}\right)
\right\},
\label{ventr:eq}
\end{align}
where
$K=1-2{R_{\rm c}}/{R}+x(1-\ln x)$, $x\equiv ({D+2R_{\rm c}})/{2R}$.
Further,
$\beta=(k_{\rm B}T)^{-1}$ is the inverse temperature, $N_{\rm trap}$
represents the number of spherically trapped ions,
$L_0$ is the
equilibrium brush height and
$R=R_{\rm c}+L_0$. 
The entropic force is given as
$F_{\rm en}(D)=-\partial V_{\rm en}(D)/\partial D$.

\begin{figure}
  \includegraphics[width=6.0cm]{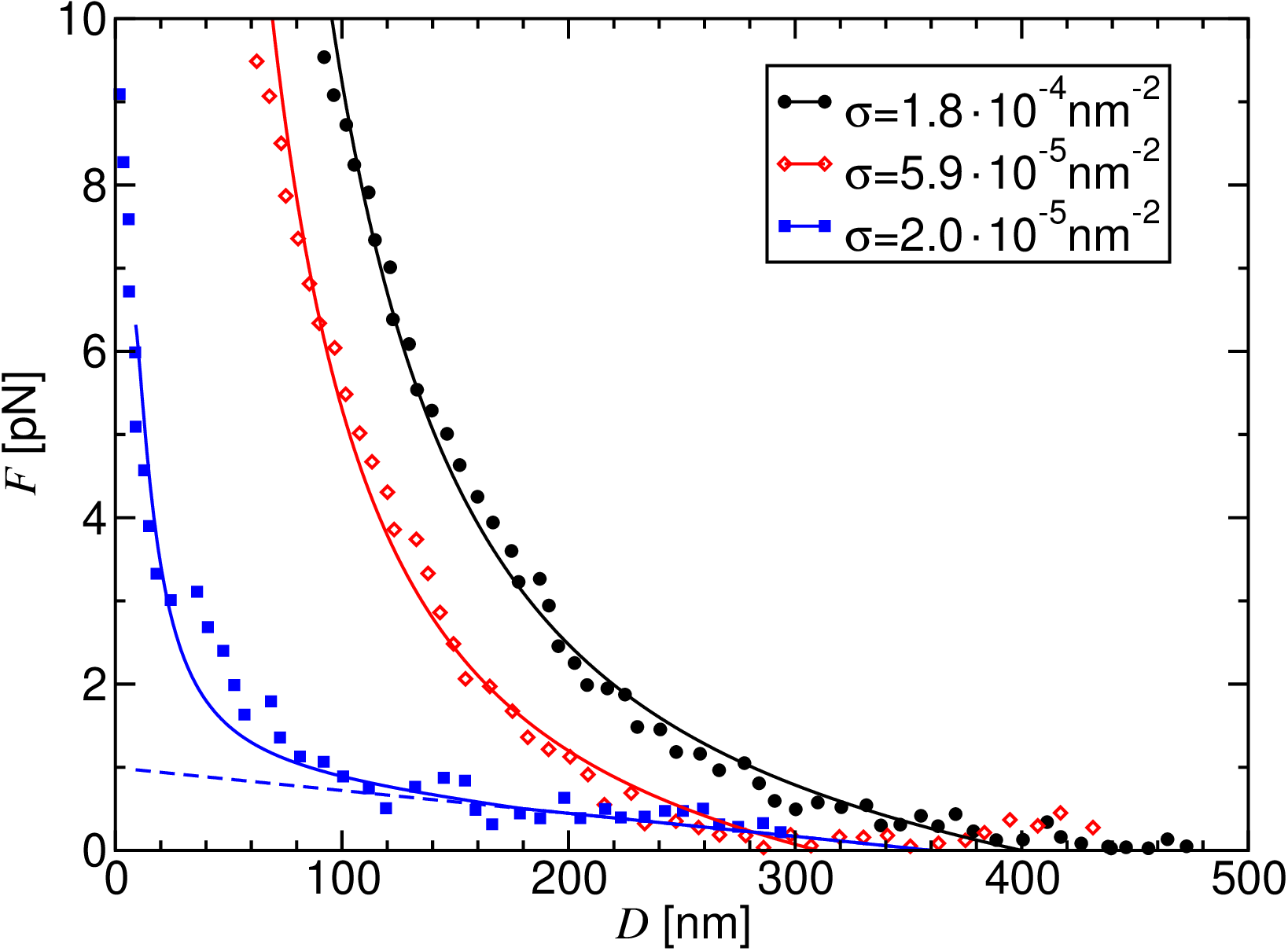}\\
  \caption{
  Effective force-distance curves for DNA-grafted colloids
  at different grafting densities $\sigma$,
  fixed molecular weight $N=1000$ bp and no added salt.
  Symbols: experiment; lines: theory. For the lowest $\sigma$,
  the solid line also includes an entropic contribution to force,
  derived from Eq.~(\ref{ventr:eq}), which is
  illustrated by the dashed
  line.}
  \label{sigma:fig}
\end{figure}

The basic assumption underlying the derivation of
Eq.~(\ref{ventr:eq}) above is that of {\it no interdigitation} between the
two brushes: as the surface-to-surface distance $D$ becomes smaller than
$2L_0$, the chains of each brush retract to the half-space
in which the respective colloidal core lies.
The experimental data at hand, however, cannot
be described by the force derived from the entropic contribution
of Eq.~(\ref{ventr:eq}), because the resulting forces have a completely
different $D$-dependence than the experimental ones (see, 
e.g., Fig.~\ref{sigma:fig}).
This is a clear indication that a different physical mechanism is at play
for the system at hand. 
The rather small grafting density of the
brushes brings about a different possibility, namely the mutual
interdigitation of the brushes up to a surface-to-surface separation
$L_0$ and the subsequent {\it compression} of the chains opposite
to the hard colloidal core for smaller distances. This mechanism
has been clearly identified and quantitatively analyzed in
Ref.~\cite{konieczny:jcp:06}, in which interactions of star-branched
polyelectrolytes with hard, planar surfaces have been discussed. Taking
into account that chains from both brushes get 
compressed
against the core of the opposite brush, the expression for the compression
contribution to the effective brush-brush interaction reads
for $d \ll D \leq L_0$ as
\begin{align}
\nonumber
\beta V_{\rm c}(D)&=
\frac{\left(Z_{\rm eff}N\right)^2\lambda_{\rm B}}{D}
\\
\times&
\left\{
2\ln\left(\frac{D}{d}\right)+
\left(\frac{D}{L_0}\right)^3\left[\ln\left(\frac{L_0}{d}\right)
-1\right]\right\}.
\label{vcompr:eq}
\end{align}

Here, the Bjerrum length $\lambda_{\rm B}=\beta e^2/\epsilon$ denotes
the distance at which the electrostatic energy equals the thermal energy 
and has the value $\lambda_{\rm B}=7.18\,{\rm {\AA}}$ for water
at $300\,{\rm K}$.
Further, $d$ is the typical diameter of individual arms of the two 
interacting brushes, having for DNA the value $d = 18\,{\rm \AA}$ 
\cite{harreis:prl:02,kornyshev:rmp:07}. Again, the compression contribution 
to the force is given as
$F_{\rm c}(D) = - \partial V_{\rm c}(D)/\partial D$ and it can be easily
checked that $F_{\rm c}(D)$ vanishes at $D=L_0$. 
Contrary to the entropic contribution, which sets in when the coronae overlap,
i.e., at $D=2L_0$, the compression contribution requires that
the grafted chains of one brush touch the core colloid of the
other and thus it is nonvanishing in the range $D \leq L_0$.

The brush height $L_0$ used in the theoretical calculation
of the forces is read off from experiment and it is thus no free fit parameter.
The `effective ion valency' $Z_{\rm eff}$ appearing in Eq.~(\ref{vcompr:eq}) 
is treated as a fit parameter to the experimental data; nevertheless, it is 
constrained by certain physical considerations based on known facts on the 
propensity of DNA strands to adsorb and strongly condense counterions on their
grooves. Indeed, as it has been shown experimentally 
\cite{bloomfield:cosb:96,kornyshev:prl:99}, DNA can strongly condense about 
90\% of counterions, which already sets a rough upper limit of $0.1$ for 
this effective valency; additional factors, such as weak adsorption of salt 
counterions inside the brush, are expected to push $Z_{\rm eff}$ to even lower 
values. 
All parameter combinations for the various DNA-grafted colloids 
are shown in Table I.

\begin{table}
\caption{The physical parameters of the employed DNA-grafted
colloids and the effective valency $Z_{\rm eff}$
employed in the theoretical modeling of each system.}
\label{table:tab}
  \begin{tabular}[t]{c|c|c|c|c|c}
  \hline\hline
Figure & $\sigma$ [${\rm nm^{-2}}$] & Base pairs &
$I$ [${\rm mM}]$ &
$L_0$ [${\rm nm}$] & $Z_{\rm eff}$ \\
    \hline
      \multirow{4}{*}{\ref{bp:fig}} & \multirow{4}{*}{$8.2\cdot 10^{-5}$} 
      & 250  & \multirow{4}{*}{0.0} & 100 & 0.227\footnote{Except for this
      value for $Z_{\rm eff}$, which is still of the order 0.1,
      all other lie below the threshold 0.1 mentioned in the text.
      Note, however, that here $L_0$ deviates from linear scaling with
      $N$, as mentioned before, and this fact may affect the precise
      $Z_{\rm eff}$-value.} \\
      & & 500  & & 130 & 0.115 \\
      & & 750  & & 180 & 0.104 \\
      & & 1000 & & 280 & 0.091\footnote{Here, different colloids were
      used than in the cases marked with the
      superscript $^c$ below, which explains
      the deviation between the corresponding 
      experimental results and the concomitant difference in 
      the $Z_{\rm eff}$-values.}\\
      \hline
      \multirow{5}{*}{\ref{salt:fig}(a)} & \multirow{5}{*}{$8.2\cdot 10^{-5}$} 
      & \multirow{5}{*}{1000} & 0.0 & 280 & 0.081\footnote{In these two cases,
      the same colloids were used, 
      yet the measured forces show
      minimal differences set by experimental accuracy. The 
      theoretical values of $Z_{\rm eff}$ are then slightly
      different between these two cases.} \\
      & & & 0.1   & 150 & 0.088 \\
      & & & 3.3   & 110 & 0.078 \\
      & & & 10.0  & 90  & 0.061 \\
      & & & 100.0 & 40  & 0.018 \\
      \hline
      \multirow{5}{*}{\ref{salt:fig}(b)} & \multirow{5}{*}{$8.2\cdot 10^{-5}$} 
      & \multirow{5}{*}{1000} & 0.03 & 180 & 0.079 \\
      & & & 0.3   & 130 & 0.077 \\
      & & & 3.0   & 100 & 0.050 \\
      & & & 9.9   & 70  & 0.045 \\
      & & & 30.0  & 40  & 0.036 \\
      \hline
      \multirow{3}{*}{\ref{salt:fig}(c)} & \multirow{3}{*}{$8.2\cdot 10^{-5}$} 
      & \multirow{3}{*}{1000} & 0.0  & 280  & 0.078$^c$ \\
      & & & 0.6  & 190  & 0.068 \\
      & & & 6.0  & 70   & 0.067 \\
      \hline
      \multirow{3}{*}{\ref{sigma:fig}} & $1.8\cdot 10^{-4}$ 
      & \multirow{3}{*}{1000} & \multirow{3}{*}{0.0}  & 400  & 0.073 \\
      & $5.9\cdot 10^{-5}$ & & & 310  & 0.056 \\
      & $2.0\cdot 10^{-5}$ & & & 180  & 0.011 \\
    \hline
    \hline
    \end{tabular}
\end{table}

The experimental forces shown in Figs.~\ref{bp:fig} and \ref{salt:fig}
are very well described by the compression force; there are still
`tails' for $D > L_0$ that can be discerned, and which correspond to
small contributions from the entropic force, Eq.~(\ref{ventr:eq}).
The purpose of this work lies in the understanding of the compression-induced
forces, thus we have not attempted a detailed description of the tails,
especially since the magnitudes of the resulting forces there 
lie at the limit of experimental accuracy. We show nevertheless 
for the lowest grafting density in Fig.~\ref{sigma:fig} a typical example 
of the combination of compression and entropic contributions that results 
into an excellent description of the forces.
Here, a number of $N_{\rm trap} = 1100$ spherically trapped
counterions was employed, which is in good agreement with the
simulation results of Ref.\ \cite{fazli:epl:06}. There, planar
PE-brushes were simulated and for a grafting density very close
to the lowest one in Fig.~\ref{sigma:fig}, it was found that about
10\% of the counterions are {\it outside} the brush. As the vast
majority of the remaining 90\% is Manning-condensed on the rods,
a very small relative number of about 1000 spherically condensed ones, 
results, in agreement with the value of $N_{\rm trap}$ mentioned above.
For the higher grafting densities shown in Fig.~\ref{sigma:fig},
the entropic contribution seems to be negligible.

Summarizing the results shown in Figs.~\ref{bp:fig}, \ref{salt:fig}, and 
\ref{sigma:fig}, it can be surmised that the compression force resulting from 
Eq.~(\ref{vcompr:eq}) yields a very good description of a large
variety of experimental data. The effective valency $Z_{\rm eff}$
from Table I always lies in the physically expected region and
shows the expected dependence on salinity, decreasing with ionic
strength $I$. Note that already the fact that the resulting forces
from theory lie at the pN-domain is a nontrivial feature, in view of
the fact that quantities of vastly different order of magnitude in SI 
(Bjerrum length, Boltzmann constant, brush height and DNA-diameter)
are involved in determining its numerical value. 

We have measured and theoretically described the forces between
spherical DNA-brushes with low grafting density. 
The physical system at hand provides a convincing verification of the
importance of the PE-compression mechanism \cite{konieczny:jcp:06},
in sharp contrast to most
hitherto studied systems, which were dominated by counterion
entropy. Therefore, it has been demonstrated that the present systems 
are colloids
whose effective interaction is short-range, i.e., tunable in
terms of its extension and strength by changing the number of base
pairs involved, the ionic strength and grafting density.
The quantitative characteristics of the resulting effective force
are unique: whereas for {\it neutral, densely grafted brushes}
the force scales as  $F(D) \sim D^{-1}$ for $D \ll L_0$
\cite{milner:macrom:88,mewis:aiche:89}, here a dependence
$F(D) \sim D^{-2}\ln(D/d)$ for $d \ll D \ll L_0$ results.
On the other hand, for $D
\lesssim L_0$, we obtain $F(D) \sim |D - L_0|$. 
Future work should
focus on the study of concentrated solutions of such brushes, including 
crystal- and glass formation, 
and the analysis of these in terms of the effective
interaction derived in this work.

We thank Dr.~Arben~Jusufi (Princeton) for helpful discussions.
This work has been supported by the DFG. 
C.N.L.~wishes to thank
the ESI (Vienna), where parts of this work have been carried out,
for its hospitality.

\bibliographystyle{unsrt}

\begin{thebibliography}{99}

\bibitem{bolhuis:prl:94} P.~Bolhuis and D.~Frenkel, Phys.\ Rev.\ Lett.\ {\bf 72},
2211 (1994).

\bibitem{jagla:99} E.~A.~Jagla, J.\ Chem.\ Phys.\ {\bf 110}, 451 (1999).

\bibitem{norizoe:epl:05} Y.~Norizoe and T.~Kawakatsu, Europhys.\ Lett.\ {\bf 72},
583 (2005).

\bibitem{glaser:epl:07} M.~A.~Glaser {\it et al.},
Europhys.\ Lett.\ {\bf 78}, 46004 (2007).

\bibitem{pablo:pre:06} Z.~Yan {\it et al.},
Phys.\ Rev.\ E {\bf 73}, 051204 (2006).

\bibitem{emanuela:pre:05} P.~Kumar {\it et al.},
Phys.\ Rev.\ E {\bf 72}, 021501 (2005).

\bibitem{largo:pre:07} J.~Largo, P.~Tartaglia, and F.~Sciortino,
Phys.\ Rev.\ E {\bf 76}, 011402 (2007).

\bibitem{pincus:mm:1} P.~Pincus,  Macromolecules {\bf 24}, 2177 (1991).

\bibitem{borisov:EPJB:1} O.~V.~Borisov and E.~B.~Zhulina, Eur.\ Phys.\ J.\
{\bf 4}, 205 (1998).

\bibitem{mei:prl:06} Y.~Mei {\it et al.},
Phys.\ Rev.\ Lett.\ {\bf 97}, 158301 (2006).

\bibitem{jusufi:cps:04} A.~Jusufi, C.~N.~Likos, and M.~Ballauff,
Colloid Polym.\ Sci.\ {\bf 282}, 910 (2004).

\bibitem{jusufi:prl:02} A.~Jusufi, C.~N.~Likos, and H.~L{\"o}wen,
Phys.\ Rev.\ Lett.\ {\bf 88}, 018301 (2002); 
J.\ Chem.\ Phys.\ {\bf 116}, 11011 (2002).

\bibitem{kegler:prl:07} K.~Kegler, M.~Salomo, and F.~Kremer,
Phys.\ Rev.\ Lett.\ {\bf 98}, 058304 (2007).

\bibitem{rabin:91} Y.~Rabin, G.~H.~Frederickson,
and P.~Pincus, Langmuir {\bf 7}, 2428 (1991).

\bibitem{konieczny:jcp:06} M.~Konieczny and C.~N.~Likos,
J.\ Chem.\ Phys.\ {\bf 124}, 214904 (2006).

\bibitem{harreis:prl:02} H.~M.~Harreis {\it et al.},
Phys.\ Rev.\ Lett.\ {\bf 89}, 018303 (2002).

\bibitem{kornyshev:rmp:07} A.~A.~Kornyshev {\it et al.},
Rev.\ Mod.\ Phys.\ {\bf 79}, 943 (2007).

\bibitem{bloomfield:cosb:96} V.~A.~Bloomfield, Curr.\ Opin.\ Struct.\
Biol.\ {\bf 6}, 334 (1996).

\bibitem{kornyshev:prl:99} A.~A.~Kornyshev and S.~Leikin,
Phys.\ Rev.\ Lett.\ {\bf 82}, 4138 (1999).

\bibitem{fazli:epl:06} H.~Fazli {\it et al.}, 
Europhys.\ Lett.\ {\bf 73}, 429 (2006).

\bibitem{milner:macrom:88} S.~T.~Milner, T.~A.~Witten, and
M.~E.~Cates, Macromolecules {\bf 21}, 2610 (1988).

\bibitem{mewis:aiche:89} J.~Mewis {\it et al.},
AIChE.\ J.\ {\bf 35}, 415 (1989).

\end{thebibliography}

\end{document}